\documentstyle[12pt]{article}
\begin{document}
\begin{center} {\large\bf CURRENT OPERATORS IN QUANTUM FIELD THEORY
AND SUM RULES IN DEEP INELASTIC SCATTERING} \end{center}
\vskip 1em
\begin{center} {\large Felix M. Lev} \end{center}
\vskip 1em
\begin{center} {\it Laboratory of Nuclear
Problems, Joint Institute for Nuclear Research, Dubna, Moscow region
141980 Russia (E-mail:  lev@nusun.jinr.dubna.su)} \end{center}
\vskip 1em
\begin{abstract}
 It is shown that in the general case the canonical construction of the
current operators in quantum field theory does not render a bona fide
vector field since Lorentz invariance is violated by Schwinger terms. We
argue that the nonexistence of the canonical current operators for
spinor fields follows from a very simple algebraic consideration. As a
result, the well-known sum rules in deep inelastic scattering are not
substantiated.
\end{abstract}

\section{Relativistic invariance of the current operators}
\label{S1}

 In any relativistic quantum theory the system under consideration is
described by some (pseudo)unitary representation of the Poincare group.
The electromagnetic or weak current operator ${\hat J}^{\mu}(x)$ for
this system (where $\mu=0,1,2,3$ and $x$ is a point in Minkowski space)
should satisfy the following necessary conditions.

 Let ${\hat U}(a)=exp(\imath {\hat P}_{\mu}a^{\mu})$ be
the representation operator corresponding to the displacement of the origin
in spacetime translation of Minkowski
space by the four-vector $a$. Here ${\hat P}=({\hat P}^0,{\hat {\bf
P}})$ is the operator of the four-momentum, ${\hat P}^0$ is the
Hamiltonian, and ${\hat {\bf P}}$ is the operator of ordinary
momentum. Let also ${\hat U}(l)$ be the representation operator
corresponding to $l\in SL(2,C)$. Then
\begin{equation}
{\hat U}(a)^{-1}{\hat  J}^{\mu}(x){\hat U}(a)=
{\hat J}^{\mu}(x-a)
\label{1}
\end{equation}
\begin{equation}
{\hat U}(l)^{-1}{\hat J}^{\mu}(x){\hat U}(l)=L(l)^{\mu}_{\nu}
{\hat J}^{\nu}(L(l)^{-1}x)
\label{2}
\end{equation}
where $L(l)$ is the element of the Lorentz group corresponding to $l$
and a sum over repeated indices $\mu,\nu=0,1,2,3$ is assumed.

 Let ${\hat M}^{\mu\nu}$ (${\hat M}^{\mu\nu}=-{\hat M}^{\nu\mu}$) be the
representation generators of the Lorentz group.  Then, as follows from
Eq. (\ref{2}), Lorentz invariance of the current operator implies
\begin{equation}
[{\hat M}^{\mu\nu},
{\hat J}^{\rho}(x)]= -\imath\{(x^{\mu}\partial^{\nu}-x^{\nu}\partial^{\mu})
{\hat J}^{\rho}(x)+g^{\mu\rho}{\hat J}^{\nu}(x)-g^{\nu\rho}
{\hat J}^{\mu}(x)\}
\label{3}
\end{equation}
where $g^{\mu\nu}$ is the metric tensor in Minkowski space.

 The operators ${\hat P}^{\mu},{\hat M}^{\mu\nu}$ act in the
scattering space of the system under consideration. In QED the electrons,
positrons and photons are the fundamental particles, and the scattering
space is the space of these almost free particles ("in" or "out" space).
Therefore it is sufficient to deal only with ${\hat P}_{ex}^{\mu},
{\hat M}_{ex}^{\mu\nu}$ where "ex" stands either for "in" or "out".
However in QCD the scattering space by no means can be considered as a
space of almost free fundamental particles --- quarks and gluons. For
example, even if the scattering space consists of one particle (say the
nucleon), this particle is the bound state of quarks and gluons, and the
 operators ${\hat P}^{\mu},{\hat M}^{\mu\nu}$ considerably differ from
the corresponding free operators $P^{\mu},M^{\mu\nu}$. It is well-known
that perturbation theory does not apply to bound states and therefore
${\hat P}^{\mu}$ and ${\hat M}^{\mu\nu}$ cannot be determined in the
framework of perturbation theory. For these reasons we will be interested
in cases when the representation operators in Eqs. (\ref{1}) and (\ref{2})
correspond to the full generators ${\hat P}^{\mu},{\hat M}^{\mu\nu}$.

 Strictly speaking, the notion of current is not necessary if the theory
is complete. For example, in QED there exist unambiguous prescriptions
for calculating the elements of the S-matrix to any desired order of
perturbation theory and this is all we need. It is believed that this
notion is useful for describing the electromagnetic or weak properties
of strongly interacted systems. It is sufficient to know the matrix
elements $\langle \beta|{\hat J}^{\mu}(x)|\alpha \rangle$ of the
operator ${\hat J}^{\mu}(x)$ between the (generalized) eigenstates of the
operator ${\hat P}^{\mu}$ such that ${\hat P}^{\mu}|\alpha\rangle =
P_{\alpha}^{\mu}|\alpha\rangle$, ${\hat P}^{\mu}|\beta\rangle=
P_{\beta}^{\mu}|\beta\rangle$. It is usually assumed that, as a
consequence of Eq. (\ref{1})
\begin{equation}
\langle \beta|{\hat J}^{\mu}(x)|\alpha \rangle=exp[\imath(P_{\beta}^{\nu}-
P_{\alpha}^{\nu})x_{\nu}] \langle \beta|{\hat J}^{\mu}|\alpha \rangle
\label{4}
\end{equation}
where formally ${\hat J}^{\mu}\equiv {\hat J}^{\mu}(0)$. Therefore
in the absence of a complete theory we can consider the less fundamental
problem of investigating the properties of the operator ${\hat J}^{\mu}$.
{}From the mathematical point of view this implies that we treat
${\hat J}^{\mu}(x)$ not as a four-dimensional operator distribution, but
as a "nonlocal" operator satisfying the condition
\begin{equation}
{\hat J}^{\mu}(x)=exp(\imath {\hat P}x){\hat J}^{\mu}
exp(-\imath {\hat P}x)
\label{5}
\end{equation}

 The standpoint that the current operator should not be treated on the
same footing as the fundamental local fields is advocated by several
authors in their investigations on current algebra (see, for example,
Ref. \cite{AFFR}). One of the arguments is that, for example, the
canonical current operator in QED is given by \cite{AB}
\begin{equation}
{\hat J}^{\mu}(x)={\cal N} \{{\hat {\bar \psi}}(x)\gamma^{\mu}{\hat
\psi}(x)\}=\frac{1}{2}[{\hat {\bar \psi}}(x),\gamma^{\mu}{\hat \psi}(x)]
\label{6}
\end{equation}
(where ${\cal N}$ stands for the normal product and ${\hat \psi}(x)$ is
the Heisenberg operator of the Dirac field), but this expression is not a
well-definition of a local operator. Indeed, Eq. (\ref{6}) involves the
product of two local field operators at coinciding points, i.e.
${\hat J}^{\mu}(x)$ is a composite operator. The problem of the correct
definition of composite operators is the difficult problem of quantum
field theory which so far has been solved only for a few models in the
framework of renormalized perturbation theory (see, for example, Refs.
\cite{Zim,Zav}) while the unambiguous way of calculating such operators
beyond perturbation theory is not known.

 It is well-known (see, for example, Ref. \cite{J1}) that it is possible
to add to the current operator the term $\partial_{\nu}X^{\mu\nu}(x)$
where $X^{\mu\nu}(x)$ is some operator antisymmetric in $\mu$ and $\nu$.
However it is usually believed \cite{J1} that the electromagnetic and
weak current operators of a strongly interacted system are given by the
canonical quark currents the form of which is similar to that in Eq.
(\ref{6}).

 We will not insist on the interpretation of the current operator
according to Eq. (\ref{5}) and will not use this expression in the
derivation of the formulas, but in some cases the notion of
${\hat J}^{\mu}$
makes it possible to explain the essence of the situation clearly.
A useful heuristic expressions which follows from Eqs. (\ref{3}) and
(\ref{5}) is
\begin{equation}
[{\hat M}^{\mu\nu},
{\hat J}^{\rho}]= -\imath (g^{\mu\rho}{\hat J}^{\nu}-g^{\nu\rho}
{\hat J}^{\mu})
\label{7}
\end{equation}

\section{Canonical quantization and the forms of relativistic
dynamics}
\label{S2}

 In the standard formulation of quantum field theory the operators
${\hat P}_{\mu},{\hat M}_{\mu\nu}$ are given by
\begin{equation}
{\hat P}_{\mu}=\int\nolimits {\hat T}_{\mu}^{\nu}(x)d\sigma_{\nu}(x),\quad
{\hat M}_{\mu\nu}=\int\nolimits {\hat M}_{\mu\nu}^{\rho}(x)d\sigma_{\rho}(x)
\label{8}
\end{equation}
where ${\hat T}_{\mu}^{\nu}(x)$ and ${\hat M}_{\mu\nu}^{\rho}(x)$ are the
energy-momentum and angular momentum tensors and
$d\sigma_{\mu}(x)=\lambda_{\mu}\delta(\lambda x-\tau)d^4x$ is
the volume element of the space-like hypersurface defined by the time-like
vector $\lambda \quad (\lambda^2=1)$ and the evolution parameter $\tau$.
In turn, these tensors are fully defined by the classical Lagrangian
and the canonical commutation relations on the hypersurface
$\sigma_{\mu}(x)$. In this connection we note that in the canonical
formalism the quantum fields are supposed to be distributions only
relative the three-dimensional variable characterizing the points of
$\sigma_{\mu}(x)$ while the dependence on the variable describing the
distance from $\sigma_{\mu}(x)$ is usual \cite{BLOT}.

 In spinor QED we define $V(x)=-L_{int}(x)=e{\hat J}^{\mu}(x)
{\hat A}_{\mu}(x)$, where $L_{int}(x)$ is the quantum interaction
Lagrangian, $e$ is the (bare) electron charge and ${\hat A}_{\mu}(x)$ is
the operator of the Maxwell field (let us note that if
${\hat J}^{\mu}(x)$ is treated as a composite operator then the product
of the operators entering into $V(x)$ should be correctly defined).

 At this stage it is not necessary to require that ${\hat J}^{\mu}(x)$ is
given by Eq. (\ref{6}), but the key assumption in the canonical
formulation of QED is that ${\hat J}^{\mu}(x)$ is
constructed only from ${\hat \psi}(x)$ (i.e. there is no dependence on
${\hat A}_{\mu}(x)$ and the derivatives of the fields ${\hat A}_{\mu}(x)$
and ${\hat \psi}(x)$). Then the canonical result derived in several
well-known textbooks and monographs (see, for example, Refs.
\cite{AB}) is
\begin{equation}
{\hat P}^{\mu}=P^{\mu}+\lambda^{\mu}\int\nolimits V(x)
\delta(\lambda x-\tau)d^4x
\label{9}
\end{equation}
\begin{equation}
{\hat M}^{\mu\nu}=M^{\mu\nu}+\int\nolimits V(x)
(x^{\nu}\lambda^{\mu}-x^{\mu}\lambda^{\nu})
\delta(\lambda x-\tau)d^4x
\label{10}
\end{equation}
 It is important to note that if $A^{\mu}(x)$, $J^{\mu}(x)$ and
$\psi(x)$ are the corresponding free operators then
${\hat A}^{\mu}(x)=A^{\mu}(x)$, ${\hat J}^{\mu}(x)=J^{\mu}(x)$ and
${\hat \psi}(x)=\psi(x)$ if $x\in \sigma_{\mu}(x)$.

  As pointed out by Dirac \cite{Dir}, any physical system can be
described in different forms of relativistic dynamics. By definition,
the description in the point form implies that the operators
${\hat U}(l)$ are the same as for noninteracting particles, i.e.
${\hat U}(l)=U(l)$ and ${\hat M}^{\mu\nu}=M^{\mu\nu}$, and thus
interaction terms can be present only in the four-momentum operators
${\hat P}$ (i.e. in the general case ${\hat P}^{\mu}\neq P^{\mu}$ for
all $\mu$). The description in the instant form implies that the
operators of ordinary momentum and angular momentum do not depend on
interactions, i.e. ${\hat {\bf P}}={\bf P}$, ${\hat {\bf M}}={\bf M}$
$({\hat {\bf M}}=({\hat M}^{23},{\hat M}^{31},{\hat M}^{12}))$, and
therefore interaction terms may be present only in ${\hat P}^0$ and the
generators of the Lorentz boosts ${\hat {\bf N}}=({\hat M}^{01},
{\hat M}^{02},{\hat M}^{03})$. In the front form with the marked $z$
axis we introduce the + and - components of the four-vectors as $x^+=
(x^0+x^z)/\sqrt{2}$, $x^-=(x^0-x^z)/\sqrt{2}$. Then we require that
the operators ${\hat P}^+,{\hat P}^j,{\hat M}^{12},{\hat M}^{+-},
{\hat M}^{+j}$ $(j=1,2)$ are the same as the corresponding free
operators, and therefore interaction terms may be present only in the
operators ${\hat M}^{-j}$ and ${\hat P}^-$.

 In quantum field theory the form of dynamics depends on the choice of
the hypersurface $\sigma_{\mu}(x)$. The representation generators of
the subgroup which leaves this hypersurface invariant are free since
the transformations from this subgroup do not involve dynamics. Therefore
it is reasonable to expect that Eqs. (\ref{9}) and (\ref{10}) give the
most general form of the Poincare group representation generators in
quantum field theory if the fields are quantized on the hypersurface
$\sigma_{\mu}(x)$, but in the general case the relation between $V(x)$
and $L_{int}(x)$ is not so simple as in QED. The fact that the operators
$V(x)$ in Eqs. (\ref{9}) and (\ref{10}) are the same follows from the
commutation relations between ${\hat P}^{\mu}$ and ${\hat M}^{\mu\nu}$.

 The most often considered case is $\tau =0$, $\lambda =(1,0,0,0)$. Then
$\delta(\lambda x-\tau)d^4x=d^3{\bf x}$ and the integration in Eqs.
(\ref{9}) and (\ref{10}) is taken over the hyperplane $x^0=0$. Therefore,
as follows from these expressions, ${\hat {\bf P}}={\bf P}$ and
${\hat {\bf M}}={\bf M}$. Hence such a choice of $\sigma_{\mu}(x)$
leads to the instant form \cite{Dir}.

  The front form can be formally obtained from Eqs. (\ref{9}) and
(\ref{10}) as follows. Consider the vector $\lambda$ with the components
\begin{equation}
\lambda^0=\frac{1}{(1-v^2)^{1/2}},\quad \lambda^j=0,\quad
\lambda^3=-\frac{v}{(1-v^2)^{1/2}}\quad (j=1,2)
\label{11}
\end{equation}
Then taking the limit $v\rightarrow 1$ in Eqs. (\ref{9}) and
(\ref{10}) we get
\begin{eqnarray}
&&{\hat P}^{\mu}=P^{\mu}+\omega^{\mu}\int\nolimits V(x)
\delta(x^+)d^4x,\nonumber\\
&&{\hat M}^{\mu\nu}=M^{\mu\nu}+\int\nolimits V(x)
(x^{\nu}\omega^{\mu}-x^{\mu}\omega^{\nu})
\delta(x^+)d^4x
\label{12}
\end{eqnarray}
where the vector $\omega$ has the components $\omega^-=1$,
$\omega^+=\omega^j=0$. It is obvious that the generators (\ref{12})) are
given in the front form and that's why Dirac \cite{Dir} related this form
to the choice of the light cone $x^+=0$.

 In Ref. \cite{Dir} the point form was related to the hypersurface
$t^2-{\bf x}^2>0,\,t>0$, but as argued by Sokolov \cite{Sok},
the point form should be related to the hyperplane orthogonal to the
four-velocity of the
system under consideration. We shall discuss this question below.

\begin{sloppypar}
\section{Incompatibility of canonical formalism with Lorentz invariance
for spinor fields}
\label{S3}
\end{sloppypar}

 A possible objection against the derivation of Eqs. (\ref{9}) and
(\ref{10}) is that the product of local operators at
one and the same value of $x$ is not a well-defined object. For
example, if $x^0=0$ then following Schwinger \cite{Schw}, instead of
Eq. (\ref{6}), one can define $J^{\mu}({\bf x})$ as the limit of the
operator
\begin{equation}
J^{\mu}({\bf x})=\frac{1}{2}[{\bar \psi}({\bf x}+\frac{{\bf l}}{2}),
\gamma^{\mu}exp(\imath e
\int_{{\bf x}-\frac{{\bf l}}{2}}^{{\bf x}+\frac{{\bf l}}{2}}
{\bf A}({\bf x}')d{\bf x}')
{\bar \psi}({\bf x}-\frac{{\bf l}}{2})]
\label{13}
\end{equation}
when ${\bf l}\rightarrow 0$, the limit should be taken only at the
final stage of calculations and in the general case the time
components of the arguments of ${\hat {\bar \psi}}$ and ${\hat \psi}$
also differ each other (the contour integral in this expression is
needed to conserve gauge invariance). Therefore there is a "hidden"
dependence of ${\hat J}^{\mu}(x)$ on ${\hat A}^{\mu}(x)$ and hence Eqs.
(\ref{9}) and (\ref{10}) are incorrect.

 However, any attempt to move apart the arguments of the ${\hat \psi}$
operators in ${\hat J}^{\mu}(x)$ immediately results in breaking of
locality. In particular, at any ${\bf l}\neq 0$ in Eq. (\ref{13}) the
Lagrangian is nonlocal. We do not think that locality is a primary
physical condition, but once the Lagrangian is nonlocal, the whole
edifice of local quantum field theory (including the canonical
Nother formalism) becomes useless. For these reason we first consider
the results which follow from the canonical formalism.

 In addition to the properties discussed above, the current operator
should also satisfy the continuity equation $\partial
{\hat J}^{\mu}(x)/\partial x^{\mu}=0$. As follows from this equation and
Eq. (\ref{1}), $[{\hat  J}^{\mu}(x),{\hat P}_{\mu}]=0$.
The canonical formalism in the instant form implies that if $x^0=0$ then
${\hat  J}^{\mu}({\bf x})=J^{\mu}({\bf x})$. Since $J^{\mu}({\bf x})$
satisfies the condition $[J^{\mu}({\bf x}),P_{\mu}]=0$, it follows from
Eq. (\ref{9}) that if ${\hat P}^{\mu}=P^{\mu}+V^{\mu}$ then the
continuity equation is satisfied only if
\begin{equation}
[V^0,J^0({\bf x})]=0
\label{14}
\end{equation}
where
\begin{equation}
V^0=\int\nolimits V({\bf x})d^3{\bf x},\quad V({\bf x}) =-e
{\bf A}({\bf x}){\bf J}({\bf x})
\label{15}
\end{equation}
We take into account the fact that the canonical quantization on the
hypersurface $x^0=0$ implies that $A^0({\bf x})=0$.

 As follows from Eqs. (\ref{1}) and (\ref{3}), the commutation relation
between the operators ${\hat M}^{0i}$ $(i=1,2,3)$ and $J^0({\bf x})=0$
should have the form
\begin{equation}
[{\hat M}^{0i},J^0({\bf x})]=-x^i[{\hat P}^0,J^0({\bf x})]-
\imath J^i({\bf x})
\label{16}
\end{equation}
Since
\begin{equation}
[M^{0i},J^0({\bf x})]=-x^i[P^0,J^0({\bf x})]-\imath J^i({\bf x})
\label{17}
\end{equation}
it follows from Eqs. (\ref{10}), (\ref{14}) and (\ref{15}) that Eq.
(\ref{16}) is satisfied if
\begin{equation}
\int\nolimits y^i {\bf A}({\bf y})[{\bf J}({\bf y}),J^0({\bf x})]
d^3{\bf y}=0
\label{18}
\end{equation}
It is well-known that if the standard equal-time commutation relations
are used naively then the commutator in Eq. ({\ref{18}) vanishes and
therefore this equation is satisfied. However when ${\bf x}\rightarrow
{\bf y}$ this commutator involves the product of four Dirac fields at
${\bf x}={\bf y}$. The famous Schwinger result \cite{Schw} is that if
the current operators in question are given by Eq. (\ref{13}) then
\begin{equation}
[J^i({\bf y}),J^0({\bf x})]=C\frac{\partial}{\partial x^i}
\delta({\bf x}-{\bf y})
\label{19}
\end{equation}
where $C$ is some (infinite) constant. Therefore Eq. ({\ref{18}) is not
satisfied and the current operator ${\hat J}^{\mu}(x)$ does not satisfy
Lorentz invariance.

 At the same time, Eq. (\ref{19}) is compatible with Eqs. (\ref{14}) and
(\ref{15}) since $div({\bf A}({\bf x}))=0$. One can also expect that the
commutator $[{\hat M}^{0i},J^k({\bf x})]$ is compatible with Eq. (\ref{3}).
This follows from the fact \cite{GJ} that if Eq. (\ref{19}) is
satisfied then the commutator $[J^i({\bf x}),J^k({\bf y})]$ does not
contain derivatives of the delta function.

 While the arguments given in
Ref. \cite{Schw} prove that the commutator in Eq. (\ref{19}) cannot vanish,
one might doubt whether the singularity of the commutator is indeed given
by the right hand side of this expression. Of course, at present any
method of calculating such a commutator is model dependent, but
the result that Eq. (\ref{16}) is incompatible with Lorentz invariance
follows in fact only from algebraic considerations. Indeed, Eqs.
(\ref{14}), (\ref{16}) and (\ref{17}) imply that if ${\hat M}^{\mu\nu}=
M^{\mu\nu}+V^{\mu\nu}$ then
\begin{equation}
[V^{0i},J^0({\bf x})]=0
\label{20}
\end{equation}

 Since $V^{0i}$ in the instant form is a nontrivial interaction
dependent operator, there is no reason to expect that it commutes with
the free operator $J^0({\bf x})$. Moreover for the analogous reason
Eq. (\ref{14}) will not be satisfied in the general case. In terms of
${\hat J}^{\mu}$ one can say that the condition
${\hat J}^{\mu}=J^{\mu}$ is incompatible with Eq. (\ref{7}) in the
instant form.

 To better understand the situation in spinor QED it is useful to
consider scalar QED \cite{J2}. The formulation of this theory can
be found, for example, in Ref. \cite{IZ}. In contrast with spinor
QED, the Schwinger term in scalar QED emerges canonically \cite{Schw,J1}.
We use $\varphi({\bf x})$ to denote the operator of the scalar complex
field at $x^0=0$. The canonical calculation yields
\begin{eqnarray}
&&{\hat J}^0({\bf x})=J^0({\bf x})=\imath[\varphi^*({\bf x}) \pi^*({\bf x})-
\pi({\bf x})\varphi({\bf x})],
\quad {\hat J}^i({\bf x})=J^i({\bf x})- \nonumber\\
&&2eA^i({\bf x})\varphi^*({\bf x})
\varphi({\bf x}),\quad J^i({\bf x})=
\imath [\varphi^*({\bf x})\cdot\partial^i
\varphi({\bf x})-\partial^i \varphi^*({\bf x})\cdot
\varphi({\bf x})]
\label{21}
\end{eqnarray}
where $\pi({\bf x})$ and $\pi^*({\bf x})$ are the operators canonically
conjugated with $\varphi({\bf x})$ and $\varphi^*({\bf x})$ respectively.
In contrast with Eq. (\ref{15}), the operator $V({\bf x})$ in scalar
QED is given by
\begin{equation}
V({\bf x}) =-e {\bf A}({\bf x}){\bf J}({\bf x})
+e^2{\bf A}({\bf x})^2\varphi^*({\bf x})\varphi({\bf x})
\label{22}
\end{equation}
However the last term in this expression does not contribute to the
commutator (\ref{16}). It is easy to demonstrate that as pointed out
in Ref. \cite{J2}, the commutation relations (\ref{3}) in scalar QED
are satisfied in the framework of the canonical formalism.
Therefore the naive treatment of the product of local operators at
coinciding points in this theory is not in conflict with the
canonical commutation relations.
The key difference between spinor QED and scalar QED is that in contrast
with spinor QED, the spatial component of the current operator is not
free if $x^0=0$ (see Eq. (\ref{21})). Just for this
reason the commutator $[{\hat M}^{0i},J^0({\bf x})]$ in scalar QED
agrees with Eq. ({\ref{3}) since the Schwinger term in this commutator
gives the interaction term in ${\hat J}^i({\bf x})$.

 Now let us return to spinor QED. As noted above, the canonical
formalism cannot be used if the current operator is considered as a
limit of the expression similar to that in Eq. (\ref{13}). In
addition, the problem exists what is the correct definition of
$V({\bf x})$ as a composite operator. One might expect that the
correct definition of $J^{\mu}({\bf x})$ and $V({\bf x})$ will result
in appearance of some additional terms in $V({\bf x})$ (and hence in $V^0$
and $V^{0i}$). However it is unlikely that this is the main reason of
the violation of Lorentz invariance. Indeed, as noted above, for only
algebraic reasons it is unlikely that both conditions (\ref{14}) and
(\ref{20}) can be simultaneously satisfied. Therefore, taking into
account the situation in scalar QED, {\it it is natural to conclude that
the main reason of the failure of canonical formalism is that either the
limit of ${\hat J}^{\mu}(x^0,{\bf x})$ when $x^0\rightarrow 0$ does not
exist or this limit is not equal to $J^{\mu}({\bf x})$} (i.e. the
relation ${\hat J}^{\mu}({\bf x})=J^{\mu}({\bf x})$ is incorrect).

 Our conclusion implies that it is insufficient to consider
${\hat J}^{\mu}({\bf x})$ as a limit of the expression in Eq. (\ref{13})
when ${\bf l}\rightarrow 0$. In turn this implies that the description
of the commutator in Eq. (\ref{19}) only by the Schwinger term (which
is a consequence of Eq. (\ref{13})) may be insufficient. The results of
several authors (see the discussion in Ref. \cite{J1}) show that if the
time and space components of the arguments $x$ and $x'$ of the fields
${\hat {\bar \psi}}(x)$ and ${\hat \psi}(x')$ defining the current
operator are different then the equal time commutators of the components
of this operator depend on the order in which the limits $x^0-x^{'0}
\rightarrow 0$ and ${\bf x}-{\bf x}'\rightarrow 0$ are calculated.

 Let us mention the mathematical fact that the commutation relations
(\ref{3}) can be formally satisfied if we assume that $V({\bf x})$ is
given by Eq. (\ref{15}), ${\hat J}^0({\bf x})=J^0({\bf x})$ but
${\hat J}^i({\bf x})=J^i({\bf x})-eCA^i({\bf x})$ where $C$ is the
same constant as in Eq. (\ref{20}). However such a form of
${\hat J}^i({\bf x})$ is not gauge invariant and therefore it cannot
be obtained from the gauge invariant expressions similar to that in
Eq. (\ref{13}). Let us also recall that in perturbative QED the
Schwinger terms do not play a role if we consider only the processes
described by connected Feynman diagrams (see, for example, Refs.
\cite{Feyn,AB,IZ}). In addition, as noted in Sec. \ref{S1}, in QED it
is sufficient to consider only commutators involving
${\hat P}_{ex}^{\mu}$ and ${\hat M}_{ex}^{\mu\nu}$. Therefore there is
no problem with Lorentz invariance of the $S$-matrix in QED. However
the above considerations are important for the problem of constructing
the current operators for strongly interacting particles (see Sec.
\ref{S4}).

\begin{sloppypar}
By analogy with Ref. \cite{Schw} it is easy to show that if $x^+=0$
then the canonical current operator in the front form $J^+(x^-,
{\bf x}_{\bot})$ (we use the subscript $\bot$ to denote the projection
of the three-dimensional vector onto the plane 12) cannot commute with
all the operators $J^i(x^-,{\bf x}_{\bot})$ $(i=-,1,2)$. As easily
follows from the continuity equation and Lorentz invariance, the
canonical operator $J^+(x^-,{\bf x}_{\bot})$ should satisfy the
relations
\begin{equation}
[V^-,J^+(x^-,{\bf x}_{\bot})]=[V^{-j},J^+(x^-,{\bf x}_{\bot})]=0
\quad (j=1,2)
\label{23}
\end{equation}
By analogy with the above consideration we conclude that these
relations cannot be simultaneously satisfied and therefore
{\it either the limit of ${\hat J}^{\mu}(x^+,x^-,{\bf x}_{\bot})$ when
$x^+\rightarrow 0$ does not exist or this limit is not equal to
$J^{\mu}(x^-,{\bf x}_{\bot})$}. Therefore the canonical light cone
quantization is incompatible with the existence of the canonical
current operator.
\end{sloppypar}

 Let us also note that if the theory should be invariant under the space
reflection or time reversal, the corresponding representation operators
in the front form ${\hat U}_P$ and ${\hat U}_T$ are necessarily
interaction dependent (this is clear, for example, from the relations
${\hat U}_PP^+{\hat U}_P^{-1}$ = ${\hat U}_TP^+{\hat U}_T^{-1}$ =
${\hat P}^-$). In terms of the operator ${\hat J}^{\mu}$ one can say
that this operator should satisfy the conditions
\begin{equation}
{\hat U}_P({\hat J}^0,{\hat {\bf J}}){\hat U}_P^{-1}=
{\hat U}_T({\hat J}^0,{\hat {\bf J}}){\hat U}_T^{-1}=
({\hat J}^0,-{\hat {\bf J}})
\label{24}
\end{equation}
Therefore it is not clear whether these conditions are compatible with
the relation ${\hat J}^{\mu}=J^{\mu}$. However this difficulty is
a consequence of the difficulty with Eq. (\ref{2}) since, as noted by
Coester \cite{Coes}, the interaction dependence of the operators
${\hat U}_P$ and ${\hat U}_T$ in the front form does not mean that there
are discrete dynamical symmetries in addition to the rotations about
transverse axes.
Indeed, the discrete transformation $P_2$ such that
$P_2\, x:= \{x^0,x_1,-x_2,x_3\}$ leaves the light front $x^+=0$ invariant.
The full space reflection $P$ is the product of $P_2$ and a rotation about
the 2-axis by $\pi$. Thus it is not an independent dynamical transformation
in addition to the rotations about transverse axes.
Similarly the transformation $TP$ leaves $x^+=0$ invariant and
$T=(TP)P_2R_2(\pi)$.

\section{Discussion}
\label{S4}

 We have shown that if the representation generators of the Poincare
group and the current operators are constructed in the framework of
the canonical formalism then in the general case the current
operators do not satisfy Lorentz invariance. One might think that
owing to some correction of the canonical formalism the current
operator will remain unchanged but the operators $V^{\mu}$ and
$V^{\mu\nu}$ will change in such a way that the conditions (\ref{14})
and (\ref{20}) in the instant form or (\ref{23}) in the front one
will be simultaneously satisfied. However, as argued in the preceding
section, there is no reason to believe that this may occur. Therefore
we conclude that in the general case the canonical current operator
does not exist and the Schwinger terms are insufficient to describe
the most singular part of the equal time commutation relations. Let us
discuss the consequences of these conclusions.

\begin{sloppypar}
 One of the problem considered in the literature (see, for example,
Ref. \cite{JL}) is the problem of constructing the
covariant $T$-product $T^*({\hat J}^{\mu}(x){\hat J}^{\nu}(y))$. It
has been shown that in the presence of Schwinger terms the standard
$T$-product $T({\hat J}^{\mu}(x){\hat J}^{\nu}(y))$ is not covariant
but it is possible to add to $T({\hat J}^{\mu}(x){\hat J}^{\nu}(y))$
a contact term (which is not equal to zero only if $x^0=y^0$) such
that the resulting $T^*$-product will be covariant. In these
investigations it was always assumed that the Lorentz invariance
condition (\ref{3}) for ${\hat J}^{\mu}(x)$ is compatible with
Schwinger terms. However in view of the above discussion the
existence of the equal time commutators cannot be guaranteed in the
general case, and if the commutators exist it is not clear whether
the main singularities are exhausted by the Schwinger terms.
\end{sloppypar}

 Let us now briefly consider the application of current algebra to
deep inelastic scattering (DIS). The hadronic tensor in DIS is
usually written as
\begin{equation}
W^{\mu\nu}=\frac{1}{4\pi}\int\nolimits e^{\imath qx} \langle N|
[{\hat J}^{\mu}(\frac{x}{2}),{\hat J}^{\nu}(-\frac{x}{2})]|N\rangle d^4x
\label{25}
\end{equation}
where $|N\rangle$ is the state of the initial nucleon and $q$ is the
4-momentum of the virtual photon, $W$ or $Z$ boson.

 The argument of many authors is that since the quantity $q$ in DIS is
very large then only the region of small $x$ contributes to the
integral (\ref{25}), and due to asymptotic freedom the current
operators in Eq. (\ref{25}) can be approximately taken free while the
corrections to such obtained expressions  are of order $\alpha_s(q^2)$
(where $\alpha_s(q^2)$ is the QCD running coupling constant).
In particular, the well-known sum rules derived from current algebra
(see, for example, Refs. \cite{Adl}) are valid. However,
as noted in Sec. \ref{S1}, the operators $V^{\mu}$ and $V^{\mu\nu}$ in
Eqs. (\ref{1}-\ref{3}) by no means can be considered as small
perturbations of the free operators $P^{\mu}$ and $M^{\mu\nu}$
respectively (see Ref. \cite{hep} for more details).

 The difficulties with equal time commutation relations were one of the
reasons for the development of the operator product expansion
(OPE) \cite{Wil}. In the framework of the OPE one can consider
$[{\hat J}^{\mu}(\frac{x}{2}),{\hat J}^{\nu}(-\frac{x}{2})]$ not only
when $x^0=0$ but also when $x^2\rightarrow 0$. Symbolically this
commutator is written as
\begin{equation}
[{\hat J}(\frac{x}{2}),{\hat J}(-\frac{x}{2})]= \sum_{i,n} C_n^i(x^2)
x_{\mu_1}\cdots x_{\mu_n} {\hat O}_i^{\mu_1\cdots \mu_n}
\label{26}
\end{equation}
where $C_n^i(x^2)$ are the Wilson coefficient $c$-number functions and
the ${\hat O}_i^{\mu_1\cdots \mu_n}$ are limits of some regular operators
${\hat O}_i^{\mu_1\cdots \mu_n}(x/2,-x/2)$ when $x\rightarrow 0$.

 The expansion (\ref{26}) has been proved only for a few models in the
framework of renormalized perturbation theory \cite{Br,Zav}.
Meanwhile this expansion is usually used when perturbation theory does
not apply, for example in DIS. Since QCD beyond perturbation theory is
very complicated, the investigations whether the OPE is valid beyond this
theory were carried out mainly in two-dimensional models (see Ref.
\cite{Nov} and references therein) and there is no agreement between
different authors whether the validity of the OPE beyond perturbation
theory can be reliably established. Besides, the authors of Ref.
\cite{Nov} did not
consider the restrictions imposed on the current operator by its
commutation relations with the representation operators of the
Poincare group. It is important to note that the Poincare group
in 1+3 spacetime is much more complicated than in 1+1 one (for example,
the Lorentz group in 1+1 spacetime is one-dimensional), and, as noted
in the preceding section, even in 1+3 spacetime the case of particles
with spin is much more complicated than the case of spinless particles.

 Let us consider DIS in the frame of reference where the $z$ component
of the momentum of the initial nucleon is positive and very large.
This frame is usually called the infinite momentum frame (IMF). It is
clear that the IMF can be obtained by choosing the four-vector $\lambda$ as
in Eq. (\ref{11}) and taking the limit $v\rightarrow 1$. Therefore one
can expect that the generators of the Poincare group in the IMF are given
in the front form. This also follows from the following facts. The
"minus" component of the four-vector $q$ in the IMF is very large and
much greater than the other components. Since $qx=q^-x^++q^+x^--
{\bf q}_{\bot}{\bf x}_{\bot}$, the integration in Eq. (\ref{25}) is
carried out over the region of very small $x^+$:
$x^+ \sim 1/q^-$
(strictly speaking such a conclusion is valid only if we consider only
Fourier transforms of smooth functions while the Wilson coefficients
are singular). Therefore the integration in Eq. (\ref{25}) involves a
small vicinity of the light cone $x^+=0$, and, as noted in Sec. \ref{S2},
the light cone quantization leads to the front form.

 Since the generators ${\hat M}^{-j}$ $(j=1,2)$ in the front form are
interaction dependent and the operators
${\hat O}_i^{\mu_1\cdots \mu_n}$ should transform as the tensors of the
$n$-th rank, these operators should properly commute with the
${\hat M}^{-j}$. Therefore, as follows from Eq. (\ref{26}), each
${\hat O}$-operator should be interaction dependent, i.e.
${\hat O}_i^{\mu_1\cdots \mu_n} \neq O_i^{\mu_1\cdots \mu_n}$ for all
$\mu_1\cdots \mu_n$ and $i$. The equality
${\hat O}_i^{\mu_1\cdots \mu_n}= O_i^{\mu_1\cdots \mu_n}$ for all
$\mu_1\cdots \mu_n$ and $i$ takes place only if the current operator
is free, i.e. in the parton model. Indeed, as shown by many authors
(see, for example, Ref. \cite{Web} and references therein), the parton
model is a consequence of the impulse approximation
(${\hat J}^{\mu}(x)=J^{\mu}(x)$) in the front form of dynamics. We see
that the parton model is incompatible with Lorentz invariance (see Ref.
\cite{hep} for details). The fact that
${\hat O}_i^{\mu_1\cdots \mu_n} \neq O_i^{\mu_1\cdots \mu_n}$
also follows from the factorization property of the OPE, according to which
the Wilson coefficients are responsible for the hard part of the
hadronic tensor and the operators ${\hat O}_i^{\mu_1\cdots \mu_n}$ are
responsible for the soft part which cannot be determined in the
framework of perturbation theory (see, for example, Ref. \cite{ER} and
references therein).

 If Eq. (\ref{26}) is valid and nothing is known about the operators
${\hat O}_i^{\mu_1\cdots \mu_n}$ then the theory makes it possible to
determine only the $q^2$ evolution of the moments of the structure
functions. However there exist a very few cases when the theory claims
that the values of the moments also can be calculated. In particular, to
lowest order in $\alpha_s(q^2)$ the theory reproduces the well-known
sum rules \cite{Adl} and then the OPE makes it possible to determine
their $q^2$ evolution.

 Let us consider the OPE to lowest order in $\alpha_s(q^2)$. Then the
conventional basis for twist two operators (see, for example, Ref.
\cite{Ynd}) is
\begin{eqnarray}
&&{\hat O}_V^{\mu_1\cdots \mu_n}=S{\cal N}\{{\hat {\bar \psi}}(0)
\gamma^{\mu_1}D^{\mu_1}\cdots D^{\mu_n}{\hat \psi}(0)\}, \nonumber\\
&&{\hat O}_A^{\mu_1\cdots \mu_n}=S{\cal N}
\{{\hat {\bar \psi}}(x)\gamma^5\gamma^{\mu}
D^{\mu_1}\cdots D^{\mu_n}{\hat \psi}(x)\}
\label{27}
\end{eqnarray}
where $D^{\mu}$ is the covariant derivative, the operator $S$ makes
the tensor traceless and for simplicity we do not write flavor
operators and color and flavor indices. In particular, it is claimed
that the operator ${\hat O}_V^{\mu}$ is equal to the electromagnetic
current operator ${\hat J}^{\mu}(0)$ (see Eq. (\ref{6})).

 The conclusion that to lowest order in $\alpha_s(q^2)$ the
basis for twist two operators is given by Eq. (\ref{27}) follows from
the assumption that since the operators ${\hat O}(x,y)$ are the basis
of the expansion of the product of two local operators depending on
$x$ and $y$ then the ${\hat O}(x,y)$ are not only regular when
$x,y\rightarrow 0$ but also bilocal in $x$ and $y$, i.e. they depend
only on fields and their derivatives at the points $x$ and $y$.
However it is obvious that gauge invariance does not allow the
operators in (\ref{27}) to be limits of some bilocal operators
(this is clear, for example, from Eq. (\ref{13})).
When we expand the product of two local operators
over some basis, we connect the points $x$ and $y$ by fields
propagating from $x$ to $y$ and therefore in the general case the
interaction dependent operator ${\hat O}(x,y)$ is nonlocal.
It is also important to note that since some of the coefficients
$C_n^i((x-y)^2)$ are singular they cannot be expanded over the
products of functions depending on $x$ and $y$.

 The operators (\ref{27}) indeed form the basis of the expansion
(\ref{26}) in the approximation when the current operator is free, i.e.
in the parton model. However in the general case the operators
${\hat O}_i^{\mu_1\cdots \mu_n}$ describe the contribution of the soft
part of the interaction between quark and gluons to the expansion
(\ref{26}), and the lowest order in $\alpha_s(q^2)$ by no means implies
that these operators have the same functional form as the free operators.

\begin{sloppypar}
As shown in the preceding section, the canonical operator
${\cal N} \{{\hat {\bar \psi}}(0)\gamma^{\mu}{\hat \psi}(0)\}$ does not
transform as a vector operator and therefore {\it if the operators in Eq.
(\ref{27}) are canonical then these operators do not transform
as the corresponding tensor operators}. This implies that the limit of
${\hat O}_i^{\mu_1\cdots \mu_n}(x/2,-x/2)$ cannot be equal to the
corresponding canonical operator in Eq. (\ref{27}).
\end{sloppypar}

 Let us suppose that the operator ${\hat O}_V^{\mu}$ depends on the
interaction between quarks and gluons at distances of order
$1/\Lambda$. The QCD running coupling constant at these distances
$\alpha_s(\Lambda^2)$ is by no means small and the perturbative
expansion over $\alpha_s(\Lambda^2)$ is meaningless. During their
propagation from $x/2$ to $-x/2$ the quark and gluon fields can produce
many virtual particles at distances $\sim 1/\Lambda$. Therefore in the
general case the limit of ${\hat O}_V^{\mu}(x/2,-x/2)$ will depend
not only on the field operators at $x=0$ but also on the integrals of
these operators over a region where these particles can be produced.
As argued in the preceding section, it is not clear whether the limit of
${\hat J}^{\mu}(x)$ when $x\rightarrow 0$ exists, and, if it exists,
it does not coincide with the canonical operator $J^{\mu}(0)$.
In addition, the analogous considerations applied to ${\hat J}^{\mu}(0)$
show that it is not clear whether it is possible to construct the
local operator ${\hat J}^{\mu}(x)$ beyond perturbation theory.
As a result, neither ${\hat O}_V^{\mu}$ nor ${\hat J}^{\mu}(0)$
can be equal to the corresponding expression in Eq.
(\ref{27}). However there is no rule which prescribes the
equality of the operators ${\hat O}_V^{\mu}$ and ${\hat J}^{\mu}(0)$.
Meanwhile the well-known sum rules \cite{Adl} (which originally
were derived from the equal time commutation relations) are based on
the equalities of some ${\hat O}$-operators and the corresponding
current operators at $x=0$. Generally speaking such equalities take
place only in the parton model and therefore we conclude that the
sum rules are not substantiated.

 In Ref. \cite{hep} a model in which the current operator explicitly
satisfies Eqs. (\ref{1}-\ref{3}) has been considered. It has been
shown that the sum rules \cite{Adl} are not satisfied in this model.
Let us stress that our considerations do not exclude a possibility
that for some reasons there may exist sum rules which are satisfied
with a good accuracy. However the statement that the sum rules
\cite{Adl} unambiguously follow from QCD is not substantiated
(it is also worth noting that these sum rules were derived when QCD
did not exist).

 In the present paper we considered the problem of constructing the
current operators in the framework of the canonical quantization of field
operators on some hypersurface. However in this case, in addition to
the difficulties discussed above, there exists the difficulty which is
ignored by many authors: once we assume that the field operators on
this hypersurface are free we immediately are in conflict with the
Haag theorem \cite{Haag,BLOT}.

Suppose we have constructed the
operators ${\hat P}^{\mu},{\hat M}^{\mu\nu}$ in the point form not
using the quantization on some hypersurface. As noted in Sec. \ref{S1},
the notion of current is not necessary, and, as follows from above
considerations, the possibility of constructing the local operator
${\hat J}^{\mu}(x)$ beyond perturbation theory is problematic. For
these reasons let us also suppose that the current operator can be
treated only according to Eq. (\ref{5}). Then
the conditions (\ref{1}-\ref{3}) are automatically satisfied if
${\hat J}^{\mu}=J^{\mu}$. The calculations of DIS on the nucleon and
deuteron assuming that this relation is valid have been carried out
in Refs. \cite{hep,hep1}. The fact that in the general case the
relation ${\hat J}^{\mu}=J^{\mu}$ is incompatible with Lorentz invariance
was pointed out in connection with the investigation of relativistic
effects in few-quark and few-nucleon systems (see, for example, Ref.
\cite{GK}).

\begin{center} {\bf Acknowledgments} \end{center}
\begin{sloppypar}
 The author is grateful to F.Coester, S.B.Gerasimov, I.L.Solovtsov and
O.Yu.Shevchenko for valuable discussions, to R.Jackiw for the proposal
to consider scalar QED and to Referee B for
constructive criticism of the original version of this paper. This work
was supported by grant No. 93-02-3754 from the Russian Foundation for
Fundamental Research.
\end{sloppypar}

\end{document}